\documentclass[12pt]{article}           
\usepackage{graphicx}
\title{\bf Are WC9 Wolf-Rayet stars in colliding-wind binaries?\footnote{Based 
on data collected at the European Southern Observatory (La Silla, Chile)} }
\author{P.~M.~Williams $^1$, K.~A.~van~der~Hucht $^{2,3}$ and G.~Rauw $^4$\\
\vspace{1cm}\\
\normalsize $^1$Institute for Astronomy, University of Edinburgh, United Kingdom\\
\normalsize $^2$SRON Netherlands Institute for Space Research, Utrecht, The Netherlands\\
\normalsize $^3$Astronomical Institute Anton Pannekoek, University of Amsterdam. 
The Netherlands\\
\normalsize $^4$University of Li\`ege, Astrophysics Institute, Belgium}
\date{\mbox{}}
\begin{document}
\maketitle
\pagestyle{empty}
%
%
\def\bull{\vrule height .9ex width .8ex depth -.1ex}
\makeatletter
\def\ps@plain{\let\@mkboth\gobbletwo
\def\@oddhead{}\def\@oddfoot{\hfil\tiny\bull\quad
``Massive Stars and High-Energy Emission in OB Associations''; JENAM
 2005, held in Li\`ege (Belgium)
\quad\bull}%
\def\@evenhead{}\let\@evenfoot\@oddfoot}
\makeatother
%
%
\def\beginrefer{\section*{References}%
\begin{quotation}\mbox{}\par}
\def\refer#1\par{{\setlength{\parindent}{-\leftmargin}\indent#1\par}}
\def\endrefer{\end{quotation}}
%
%
{\noindent\small{\bf Abstract:}
We present results from a spectroscopic search for massive companions to  
dust-making Galactic WC9 stars as a step to testing the paradigm that dust 
formation in these systems requires colliding winds to produce over densities. 
We find evidence for OB companions to the WC9 stars WR\,59 and WR\,65, but 
not WR\,121 or WR\,117. 
We identify lines of N\,{\sc iii-v} and possibly N\,{\sc ii} in the spectrum 
of WR\,88, one of the few Galactic WC9 stars which do not make circumstellar 
dust, and suggest that WR\,88 is a transitional WN--WC9 object and less evolved 
than the other WC9 stars. On the other hand, the possible identification 
of a strong emission line at 4176\AA\ in the spectrum of WR\,117 with 
Ne\,{\sc i} suggests that this star is more evolved than other WC9 stars 
studied. 
}
%
%
\section{Wolf-Rayet dust formation and binarity}

Infrared observations over the last 30+ years have shown that WC-type Wolf-Rayet 
stars make dust, but the mechanism for this is still not understood. Studies have 
shown that dust formation by WR stars requires that their winds contain regions 
of significantly higher density. These can be provided by shocks in colliding 
stellar winds (Usov 1991) if the WR stars are members of massive binary systems. 
Observational support for this process comes from the infrared light curve of the 
episodic WC7+O5 dust-maker WR\,140, which makes dust very briefly during periastron 
passage when the pre-shock wind density is highest (Williams 1999), and from the 
rotating `pinwheels' of heated dust formed by the persistent dust-makers WR\,104 
(Tuthill, Monnier \& Danchi 1999) and WR\,98a (Monnier, Tuthill \& Danchi 2000), 
which are considered to be binaries observed at low inclination angle. 
The presence of heated dust around most WC9 stars then prompts the question: 
are all dust-making WC9 stars colliding-wind binaries? As a step to answering this 
question, we observed (Williams \& van der Hucht 2000, Paper~I) a selection of 
WC9 stars with the 1.9-m telescope at the SAAO and found absorption lines 
attributable to companions to the WC9 stars in the spectra of WR\,104 and WR\,69. 
We also found spectroscopic differences between two non-dusty WC9 stars and the 
dusty stars in our sample, suggesting a compositional difference having a bearing 
on dust formation.

\begin{figure}
\centering
\includegraphics[width=11cm]{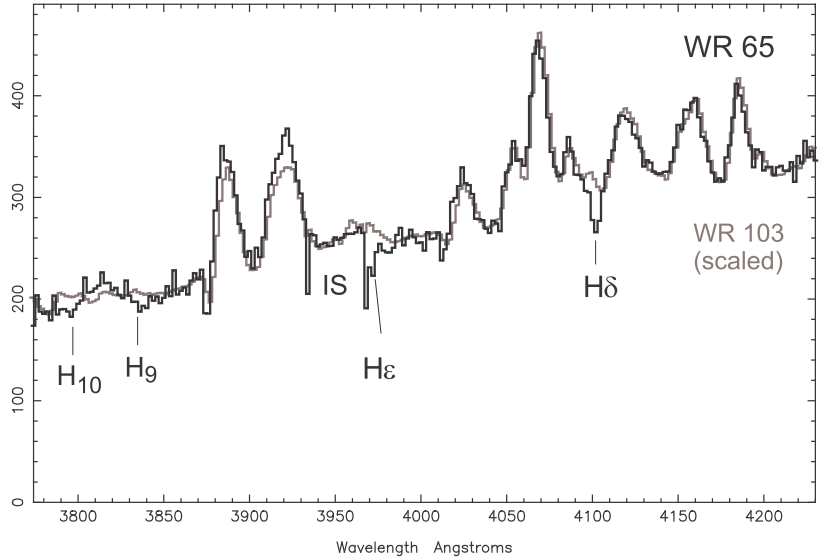}
\caption{The spectrum of WR\,65 compared with that of WR\,103 (grey), scaled 
to match that of WR\,65. Balmer lines H$_{\rm 10}$, H$_{\rm 9}$, H$\epsilon$ 
(in the wing of interstellar Ca\,{\sc ii} H) and H$\delta$ are clearly visible 
in WR\,65.}
\end{figure}

Here we report a follow-up of that study, based on medium-dispersion 
($\sim$ 1\AA) spectra observed with EMMI on the ESO NTT in 2001 June. 
We used Grating~3 at two settings, centred near 4000\AA\ (`violet') and 
4150\AA\ (`blue'), to search 
for H$\gamma$ and higher-$n$ Balmer and He\,{\sc i} absorption lines 
indicative of OB companions. For our candidates, we included the 
WC9 stars noted as having diluted emission-line spectra in the VIIth 
Catalogue (van der Hucht 2001). Because H$\gamma$ and H$\delta$ lie close 
to emission lines, we found that H$\epsilon$ and H$_{\rm 9}$, which are 
located in clearer regions of the spectrum, were good diagnostics despite 
their relative weakness. 

We discovered Balmer absorption lines in our EMMI spectra of WR\,59 and 
WR\,65, but not in those of WR\,121 or WR\,117. 
The `violet' spectrum of WR\,65 is shown in Fig.\,1, where it is 
compared with the spectrum of WR\,103, a well-observed, `typical' WC9 
star. The `blue' spectrum showed H$\gamma$ absorption. Our 
spectra of WR\,59 are similar to those of WR\,65, but also show weak 
absorption on top of the 4025\AA\ He\,{\sc i} + He\,{\sc ii} and 
4472\AA\ He\,{\sc i} profiles, and also possibly He\,{\sc i} at 4387\AA.
From the apparent absence of He\,{\sc i} absorption in WR\,65, its 
companion may be of earlier subtype than that in WR\,59 but further 
work, using synthetic composite spectra, is needed to estimate strengths 
of He\,{\sc ii} lines to classify the companions. Together with Paper~I, 
we have spectroscopic companions to 4/11 WC9 systems observed. Our 
results suggest we should detect any OB companions at least as luminous 
as the WC9 stars. The three WC9 stars with known luminosities have 
$-4.16 \geq M_v \geq -4.97$ (VIIth Catalogue), so there could still 
be undetected main-sequence OB companions.

Confirmation that the absorption lines do come from a companion require 
observation of RV shifts attributable to orbital motion. We re-observed 
WR\,69, found to have a companion in our SAAO spectroscopy, and found 
that there was indeed a relative shift in RVs (absorption -- emission) 
between SAAO and ESO observations, making this star a prime candidate for 
an orbital analysis.

\section{Spectral diversity: an evolutionary sequence?} 

We observed in Paper I that the spectra of two WC9 stars (WR\,81 and WR\,92), 
which had never (in 20+ years of IR photometry) shown dust emission, 
differed from the other WC9 stars in having weaker O\,{\sc ii} (relative to 
C\,{\sc ii}), and stronger He\,{\sc ii} lines, suggesting a compositional 
difference. Previously, Torres \& Conti (1984) had found that the spectrum 
of the dust-free WC9 star WR\,88 differed from those of the other WC9 stars 
in having stronger He\,{\sc ii} and weaker C\,{\sc ii} lines. 
We re-observed WR\,88 with EMMI to see if its O\,{\sc ii}/C\,{\sc ii} ratio 
resembled those of WR\,81 and WR\,92 in being lower than in dust-forming 
WC9 stars. 
Our spectra of WR\,88 (the `violet' spectrum is compared with that 
of WR\,92 in Fig. 2) confirmed the strengths of the lines at 4101\AA\ and 
4200\AA\, but other He\,{\sc ii} lines, e.g. those near 3968\AA\ and 3923\AA, 
are not stronger. We deduce that most of the 4101\AA\ feature in WR\,88 
must come from another ion, for which we propose multiplet 1 of N\,{\sc iii}. 
Multiplet 6 of N\,{\sc iii} could contribute to He\,{\sc ii} line at 
4200\AA\ and we observe multiplet 17 at 4379\AA\ in our `blue' spectrum. 
The 4650\AA\ C\,{\sc iii} line is stronger and broader in WR\,88 and does 
not have the P-Cygni absorption component seen in WR\,92, possibly due to 
the strong N\,{\sc iii} 4640\AA\ multiplet. 
We confirm the identification of the 4603\AA\ line with N\,{\sc v} as we 
also see the 4619\AA\ line of the same multiplet. 
As shown in Fig. 2, WR\,88 appears to have N\,{\sc iv} 4058\AA\ blended 
with C\,{\sc iii} 4056\AA, which is much weaker in WR\,92. 

\begin{figure}
\centering
\includegraphics[width=11cm]{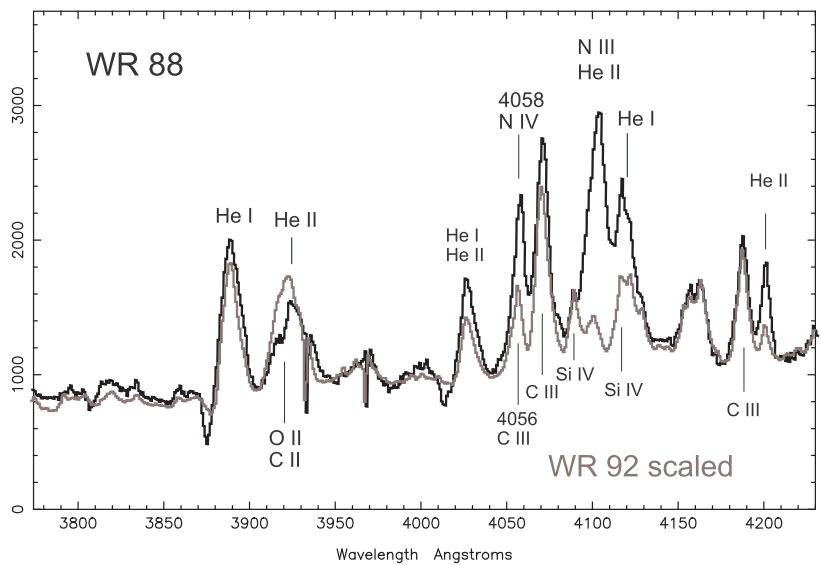}
\caption{Part of the spectrum of WR\,88, compared with that of another 
dust-free WC9 star, WR\,92 (grey), scaled to match the C\,{\sc iii} 
lines in WR\,88. From comparison of the He\,{\sc ii} lines, we believe 
most of the 4101\AA\ line comes from another ion, which we suggest is 
N\,{\sc iii}, while from comparison of the C\,{\sc iii} lines, we 
suggest that N\,{\sc iv} is a major contributor to the line at 4058\AA.}
\end{figure}

We examined the 10-\AA\ resolution spectra of WR\,88 and WR\,92 in the 
Torres \& Massey (1987) atlas and observed other differences, e.g. the 
presence in WR\,88 of the N\,{\sc iv} 5200\AA\ and 7117\AA\ and N\,{\sc v} 
4933-45\AA\ lines seen in WN stars (cf. Hamann, Koesterke \& Wessolowski 
1995). We may also be seeing some N\,{\sc ii} lines in WR\,88, but this 
needs confirmation with higher resolution spectroscopy.

We conclude that WR\,88 either has a WN companion or is of a previously 
unobserved transitional WN--WC9 type, and prefer the latter alternative 
since the N lines have comparable widths to the C lines, suggesting 
formation in the same wind. This would make WR\,88 less evolved than 
WR\,92 and WR\,81, and even less evolved than the dust-making WC9 stars.

\begin{figure}
\centering
\includegraphics[width=11cm]{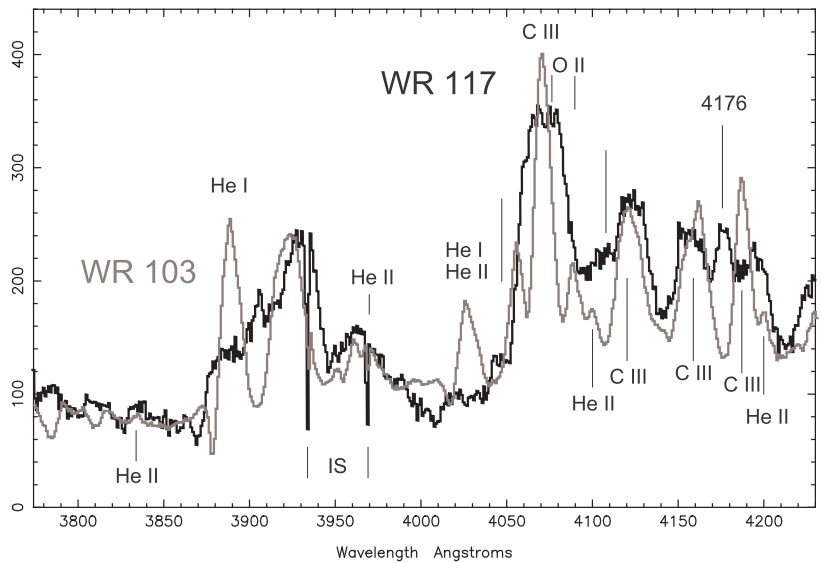}
\caption{Spectrum of WR\,117, compared with that of WR\,103 (grey). Note 
the weakness of the He\,{\sc i} lines in WR\,117, and the line at 4176\AA.} 

\end{figure}

While WR\,88, WR\,92 and WR\,81 appear to be less evolved than most of the 
WC9 stars, we found one star to be apparently more evolved: WR\,117.
We had time to observe only a `violet' spectrum (Fig. 3), which appears 
to be significantly different from that of WR\,103. The lines in WR\,117 are 
broader than in WR\,103, but comparisons with another broad-lined WC9 star, 
WR\,53 (Paper I), and the WC8 star WR\,135 (Torres \& Massey), show similar 
differences. The feature at 4176\AA\ does not appear to coincide with ions 
usually seen in WR stars, and we identify it with the 
2p$^5$($^2$P$^{\rm o}_{3/2}$)3p--2p$^5$($^2$P$^{\rm o}_{3/2}$)8d 
array of Ne\,{\sc i} (van Hoof 1999). We urgently need spectroscopy in the 
red to confirm the presence of Ne\,{\sc i}, which would make WR\,117 more 
evolved than other WC9 stars. The apparent evolutionary sequence exemplified 
by WR\,88 $\rightarrow$ (WR\,81, WR\,92) $\rightarrow$ (WR\,103 etc) 
$\rightarrow$ WR\,117 needs to be tested by full spectroscopic analyses. 

%
%
%
%
 
\beginrefer

\refer Hamann W.-R., Koesterke L., Wessolowski U., 1995, A\&A Supp. 113, 459
 
\refer Monnier J. D., Tuthill P. G., Danchi W. C., 1999, ApJ 525, L97

\refer Torres A. V., Massey P., 1987, ApJS 65, 459

\refer Torres A. V., Conti P. S., 1984, ApJ 280, 181

\refer Tuthill P. G., Monnier J. D., \& Danchi W. C., 1999, Nature, 398, 486

\refer Usov V. V., 1991, MNRAS 252, 49

\refer van der Hucht K. A., 2001. New Astron. Revs 45, 135 (VIIth Catalogue)

\refer van Hoof P., 1999. The Atomic Line List v2.04, 
{\tt http://www.pa.uky.edu/$\sim$peter/atomic/}

\refer Williams, P.M. 1999. In: K. A. van der Hucht, G., Koenigsberger \& 
P. R. J. Eenens (eds), Proc. IAU Symp.\,193, Wolf-Rayet Phenomena in Massive Stars 
and Starburst Galaxies. San Francisco, ASP. p.~267

\refer Williams P. M., van der Hucht K.A., 2000, MNRAS 314, 23 (Paper I) 

\endrefer           
\end{document}